\documentclass[pdflatex,sn-mathphys-num]{sn-jnl}
\usepackage{amsmath,amssymb,amsfonts}
\usepackage[nolist]{acronym}
\usepackage{graphicx}
\usepackage{braket}
\usepackage{multirow}

\usepackage{algorithmicx}
\usepackage{algorithm2e}
\usepackage{listings}

\begin{acronym}[XXX]
    \acro{nisq}[NISQ]{Noisy Intermediate-Scale Quantum computers}
    \acro{dd}[DD]{Decision Diagram}
    \acro{qmdd}[QMDD]{Quantum Multiple-Valued Decision Diagram}
    \acro{tdd}[TDD]{Tensor Decision Diagram}
    \acro{ftdd}[FTDD]{Fast Tensor Decision Diagram}
    \acro{robdd}[ROBDD]{Reduced Ordered Binary Decision Diagram}
    \acro{tn}[TN]{Tensor Network}
    \acro{bdd}[BDD]{Binary Decision Diagrams}
    \acro{qc}[QC]{Quantum Computing}
    \acro{hpc}[HPC]{High Performance Computing}
    \acro{nr}[\texttt{NR}]{\texttt{Node-Reuse}}
    \acro{nrtr}[\texttt{NRTR}]{\texttt{Node-Reuse and Table-Resize}}
\end{acronym}

\newcommand{\C}{\mathbb{C}}
\newcommand{\cotengra}{\texttt{cotengra} }

\newcommand{\seq}{\texttt{sequential} }

\newcommand{\alp}{\texttt{alphanumeric} }
\newcommand{\rcm}{\texttt{RCM} }
\newcommand{\mpath}{\texttt{path} }

\newcommand{\ct}{\texttt{ComputeTable} }
\newcommand{\ut}{\texttt{UniqueTable} }
\title{Optimizing Memory Efficiency and Index Ordering to Simulate Quantum Circuits Using Tensor Decision Diagrams}

\abstract{
Combining Tensor Networks (TNs) and Decision Diagrams (DDs) provides a high-performance framework for the exact simulation of quantum circuits on classical computers by exploiting structural redundancies and topological entanglement. However, improving the scalability of hybrid tools such as the Fast Tensor Decision Diagram (FTDD) requires addressing strict memory growth constraints and complex node management during tensor contractions. In this paper, we propose a hardware-aware architectural optimization of the FTDD framework, with two main contributions: first, we overhaul the internal memory management through a fixed-footprint memory model, strict node lifecycle tracking, and a systematic evaluation of table-sizing policies (exponential, static, and hybrid) to reduce allocation overhead and control memory growth; second, we introduce Path, a new index-ordering heuristic guided by the contraction path, and conduct a comprehensive study of variable index-ordering strategies, an important factor for DD compression. By comparing the original alphanumeric ordering, RCM, and our Path heuristic across diverse circuit topologies, we show that index permutation strongly affects node sharing and diagram density. Experimental evaluation confirms that our optimized FTDD engine bounds memory consumption under structured quantum workloads, enabling stable execution of circuits such as QFT with up to 100 qubits. Furthermore, we systematically characterize execution time, memory footprint, and topological trade-offs across diverse quantum benchmarks.
}

\begin{document}

\author*[1]{\fnm{Vicente} \sur{Lopez-Oliva}}\email{voliva@uji.es}

\author[1]{\fnm{Jose M.} \sur{Badia}}\email{badia@uji.es}

\author[1]{\fnm{Maribel} \sur{Castillo}}\email{castillo@uji.es}

\affil*[1]{\orgdiv{Dept. of Comp. Sci. and Eng.}, \orgname{Universitat Jaume I de Castelló}, \orgaddress{\street{Avda. Sos Baynat, s/n}, \city{Castelló de la Plana}, \postcode{12071}, \state{Castelló}, \country{Spain}}}

\maketitle

\section{Introduction}
The field of \ac{qc} has attracted significant attention due to its transformative potential across various scientific domains~\cite{YUA20}. However, the capabilities of current quantum hardware remain severely constrained by the limitations of the \ac{nisq} era~\cite{Pre18}, most notably a restricted qubit count and high susceptibility to environmental noise. These architectural imperfections hinder the reliable execution of advanced quantum algorithms, rendering classical simulation an indispensable tool for designing, validating, and optimizing quantum algorithms. Thus, the transition toward practical \ac{qc} relies heavily on our capacity to validate, benchmark, and verify quantum algorithms on classical \ac{hpc} infrastructures. State-of-the-art simulators broadly follow distinct paradigms, each with clear trade-offs: full state-vector backends track all $2^N$ complex amplitudes but hit a strict memory wall at $N \approx 50$ qubits~\cite{CMA25}; \acp{tn} excel at modeling localized topological entanglement through tensor contractions~\cite{CSW97}, but scale poorly with circuit depth. Even when supercomputers are deployed with thousands of nodes, simulation remains limited to specific circuits with only a few hundred qubits. This leaves a critical gap in benchmarking tools capable of evaluating systematic coherent noise at the scale of thousands of qubits.

As traditional simulation approaches based on explicit matrix and vector multiplications face critical scalability challenges, the use of \ac{dd} were proposed to reduce the memory needed for simulating quantum circuits~\cite{HZN21}. While \acp{tn} excel at tracking topological entanglement, \acp{dd} provide exceptional structural compression by exploiting regularities, symmetries, and redundancies within quantum states and operators~\cite{ZuW18}.
This advantage comes at a cost, as it is notoriously difficult to parallelize efficiently except in very specific scenarios~\cite{HZW20}. Consequently, simulators based on \acp{dd} are generally sequential in nature.

To combine the strengths of tensor-based connectivity representation and \ac{dd} state compression, hybrid engines such as the \ac{ftdd} framework~\cite{ZSK24} have been developed. This unique combination provides a promising path toward scaling exact quantum circuit simulation beyond standard state-vector bounds. Nevertheless, the scalability of the current implementation is limited by significant memory-management and computational bottlenecks. During complex tensor contractions, the internal representation may grow rapidly, leading to excessive memory consumption, inefficient data reuse, and substantial management overhead. These limitations become particularly critical for deep, highly entangled, or structurally complex quantum circuits, where the simulator may exhaust the available physical memory before completing the simulation.

This paper addresses these limitations through a systematic, hardware-aware optimization of the \ac{ftdd} framework\footnote{https://github.com/voliva-esp/FETDD}. To this end, this work presents a two-fold contribution: first, an architectural and memory-level optimization of the simulator to enhance its robustness and performance; second, a comprehensive study on the critical role of index-ordering strategies in tensor contraction efficiency. Rather than treating these aspects as independent problems, the proposed approach jointly considers memory organization, internal data-structure management, allocation policies, and index ordering to effectively control memory growth and evaluate the trade-offs between memory consumption and execution time across diverse quantum circuit topologies.
The core contributions of this work are the following:

\begin{enumerate}
\item \textbf{A comprehensive metric-driven experimental evaluation:} We establish a rigorous benchmark suite comprising a wide and diverse range of quantum circuits (e.g., QFT, Sycamore-like layers, and algorithmic benchmarks). These circuits are chosen using structural metrics—such as entanglement entropy, gate density, and depth to evaluate the multidimensional trade-offs between peak memory consumption and execution time.

\item \textbf{A bounded and hardware-aware memory-management strategy:} We redesign the core memory mechanisms of the \ac{ftdd} engine by implementing custom pool allocators, selecting caching operations, and optimized node reuse policies. This drastically mitigates allocation overhead and prevents memory exhaustion during dense tensor contractions.

\item \textbf{A systematic analysis of memory-allocation policies:} We evaluate alternative allocation and sizing strategies and characterize their impact on memory consumption, garbage-collection overhead, data-access efficiency, and overall execution time.

\item \textbf{A new index ordering strategy:} We propose a new strategy to order the indices called \mpath based on the \textit{cut-width} minimization problem for the index interaction graph, guided by the contraction path. This new order minimizes the distance between the indices involved in each contraction.

\item \textbf{An evaluation of alternative index-ordering strategies:} We implement and assess different internal index orderings across quantum circuits with diverse structures and topologies, analysing their influence on contraction efficiency, memory requirements, and simulation performance.

\end{enumerate}

Our empirical findings demonstrate that the proposed hardware-aware memory strategy effectively limits and reduces the maximum amount of memory used. Furthermore, the results reveal that optimal index ordering can reduce the amount of time consumed in the simulation by one order of magnitude. Together, these optimizations allow the enhanced \ac{ftdd} simulator to execute previously intractable deep and highly entangled quantum circuits, providing a robust, high-performance tool for exact quantum simulation. Remarkably, despite being a sequential simulator, our implementation successfully simulates quantum circuits of up to 100 qubits on a high-memory server. This positions the proposed framework competitively against state-of-the-art parallel simulators running on top-tier supercomputers for specific circuit families~\cite{CMA25}.

The paper is structured as follows: Section~\ref{sec:rw} briefly reviews some related work. Section~\ref{sec:back} summaries the theoretical background of our work, including basic aspects of quantum computation, tensor networks, and decision diagrams. In section~\ref{sec:memory}, we detail the improvements regarding the memory management. The description of the alternative global index ordering introduced in this tool is summarized in section~\ref{sec:order}. In section~\ref{sec:res}, we provide a detailed description of the experiments conducted and discuss the results. Finally, Section~\ref{sec:con} presents the conclusions of the study and suggests directions for future research.

\section{Related Work}
\label{sec:rw}

The classical simulation of quantum circuits represents a fundamental benchmark for demonstrating quantum advantage and validating near-term quantum algorithms. Over the past decades, more than 100 different simulators with different simulation strategies have been developed~\cite{Qua23}. The three primary paradigms are: full-statevector and density-matrix \ac{hpc} frameworks, \ac{tn} contraction schemes, and \ac{dd}-based state compression.

The foundational motivation for quantum simulation stems from Feynman~\cite{Fey82}, who highlighted the exponential memory barrier inherent in classically modeling quantum mechanical systems. To enable algorithm design and hardware calibration, open-source simulation frameworks have become indispensable tools for the \ac{qc} community. Among these, Qiskit Aer~\cite{Qis23} has emerged as one of the most widely adopted frameworks, providing versatile state-vector, unitary, and density-matrix backends integrated directly with physical hardware noise models. Similarly, open-source platforms such as Google's Cirq/qsim~\cite{OTC20} and Qibo~\cite{ERB22} offer flexible interfaces for gate-level circuit execution and noisy channel emulation.

Full-statevector simulation tracks the complete $2^N$ complex amplitude vector of an $N$-qubit quantum state. To extend state-vector tracking to larger qubit registers, \ac{hpc} environments distribute the $2^N$ state vector across massively parallel clusters. Frameworks like qHiPSTER~\cite{SSA16} and the JUWELS massively parallel simulator, that was first introduced in~\cite{DMD07} and then further improved in~\cite{DJW19}. This simulator leverages petascale RAM architectures to execute deep quantum circuits. However, direct statevector simulation encounters a strict ``memory wall" at roughly 45–50 qubits~\cite{PGN17, PGN19, CMA25}, beyond which the memory required to store the $2^N$ complex amplitudes exceeds petabyte capacities regardless of raw computational power. To bypass the exponential memory cost of storing full statevectors, \ac{tn} contraction techniques represent quantum states compactly by exploiting low entanglement entropy rather than tracking the entire Hilbert space. Vidal introduced in~\cite{Vid03} Matrix Product States (MPS) and the Time-Evolving Block Decoupling (TEBD) algorithm, enabling efficient simulation for one-dimensional quantum circuits with bounded entanglement. This concept was further generalized to higher-dimensional topologies using Projected Entangled Pair States (PEPS)~\cite{VeC04} and Tree Tensor Networks (TTN)~\cite{SDV06}.

For general circuit topologies, Markov and Shi~\cite{MaS08} formalized \ac{qc} simulation as the contraction of a 3D \ac{tn}, proving that computational complexity is dictated by the tree-width of the underlying circuit graph. Modern tensor contraction engines~\cite{PCZ22} utilize advanced slicing and bond-dimension truncation techniques to simulate complex quantum circuits. 
Some of the most widely used simulators leveraging this technique include quimb, which has been extensively utilized to demonstrate large-scale \ac{tn} contraction algorithms for highly entangled circuits~\cite{GrK21}. 
Nevertheless, \ac{tn} methods suffer severe performance degradation when simulating highly entangled circuits, as high entanglement rapidly expands the required tensor rank, forcing a trade-off between exactness and computational feasibility.

\acp{dd} offer a complementary compression paradigm by exploiting algebraic redundancies, zero blocks, and identity sub-matrices within quantum operators and state vectors. Viamontes et al.~\cite{VMH03} pioneered this direction with QuIDDPro, adapting Algebraic Decision Diagrams (ADDs) to represent quantum states compactly. Later, in~\cite{MiT06} was extended this concept through Quantum Multiple-Valued Decision Diagrams (QMDDs), establishing efficient data structures for quantum logic synthesis, verification, and equivalence checking. Bridging \ac{tn} and \ac{dd}, in~\cite{HZL22} were introduced Tensor Decision Diagrams (TDDs), which replace traditional tensor storage with canonical decision tree representations to achieve aggressive node sharing. Building upon this, the Fast Tensor Decision Diagram (FTDD) framework further optimizes node reduction~\cite{ZSK24}.

\section{Background}
\label{sec:back} 

To establish a self-contained context, this section surveys the core theoretical concepts essential to our development. A more detailed analytical account of the quantum computing landscape is available in~\cite{Sut19} and~\cite{Lop22}.

\subsection{Quantum Circuits as Tensor Networks}
Analogous to the classical bit, the fundamental unit of quantum information is the qubit, formally represented as a linear superposition of the basis states, $\alpha \ket{0} + \beta\ket{1}$, where $\alpha, \beta \in \C$~\cite{Dir39}. A composite quantum system comprising $n$ qubits, denoted as $\phi_1, \dots, \phi_n$, resides within a Hilbert space of dimension $2^n$. The computational basis of this space is spanned by the set $\{0, 1\}^n$; consequently, any arbitrary quantum state $\ket{\psi}$ can be characterized as:

\begin{equation}
 \ket{\psi} = \sum_{i \in \{0, 1\}^n}\alpha_i\ket{i} \text{ with }
 \forall i \in \{0, 1\}^n, \alpha_i \in \C,
\end{equation}

\noindent where $|\alpha_i|^2$ denotes the probability of obtaining the state $\ket{i}$ upon measurement. Quantum systems are transformed via unitary operators of dimension $2^n \times 2^n$, which typically operate on specific qubit subspaces. The serial execution of these operators, $g_1, \dots, g_k$, is conventionally depicted through quantum circuits. From this perspective, a quantum algorithm is essentially a circuit that defines an overarching unitary transformation representing the algorithm's complete logical operation.

From a broader perspective, tensors extend the concept of matrices to arbitrary orders. Specifically, a rank-$r$ tensor is an object residing in the space $\mathbb{C}^{d_1 \times \dots \times d_r}$. This hierarchy classifies scalars as rank-0 tensors, vectors as rank-1, and $n \times m$ matrices as rank-2 tensors. A robust set of operations governs these structures, including the tensor product, trace, and partitioning~\cite{BrC17}. A pivotal operation in this context is contraction: the process of generating a new tensor from $R_{x, z}$ and $S_{y,z}$ by summing over shared indices, defined as follows:

\begin{equation}
 T_{x, y} = \sum_{z \in \{0, 1\}} R_{x, z} \cdot S_{y,z}
\end{equation} 

Broadly speaking, the sequential contraction of tensor pairs within a network facilitates the derivation of a single, global tensor that encapsulates the system's entire functionality. 

The quantum state $\ket{\psi}$ of an individual qubit can be formalized as a vector

\begin{equation}
 \ket{\psi} = [\alpha_0, \alpha_1] ^ T \text{, where } \alpha_0, \alpha_1 \in \C.
\end{equation}

Thus, the qubit is represented as a first-order tensor $T_q$, while the associated quantum gates are characterized as rank-2 tensors $T_{q_1, q_2}$. This mapping extends to more complex operations: a gate acting on a manifold of $n$ qubits is represented as a tensor of rank $2n$, effectively capturing the multilinear interaction between the constituent subsystems.

Quantum circuits admit a natural mapping to a single \ac{tn}, where individual gates are cast as tensors. Since the functional behavior of a circuit is traditionally captured by matrix multiplication, it can, by extension, be simulated through tensor operations. This equivalence recasts the simulation task as a \ac{tn} contraction problem, a paradigm shift that serves as the foundation for the approach introduced by Markov and Shi to optimize the classical simulation of quantum systems~\cite{MaS08}.

\subsection{Decision Diagrams}


Traditionally, the manipulation of quantum circuits is performed through matrix algebra, a method favored for its access to high-performance computational kernels. Nevertheless, symbolic representations, particularly \acp{dd}, have emerged as a powerful paradigm for exploiting structural redundancies within quantum states. Since the formalization of \acp{bdd}~\cite{BRB91}, these diagrams have been adapted to various domains~\cite{HDH24, FDL16}. By employing directed acyclic graphs, \acp{dd} effectively encapsulate the gate sequences and the resulting unitary operators, offering a versatile framework for complex tasks such as quantum system simulation and verification. In the quantum domain, the first attempt can be found in~\cite{Ran86}, using this representation to simulate quantum circuits. Since then, many tools have been developed using this type of diagrams. The most prominent implementation is the MQT (formerly JKQ) framework~\cite{ZuW18}, which utilizes \acp{qmdd}. These diagrams represent quantum gates and states by recursively partitioning matrices into sub-quadrants until reaching scalar values. While \acp{qmdd} excel at identifying identical sub-structures, their traditional implementation is somewhat rigid, as it often imposes a fixed variable ordering that may not align with the entanglement dynamics of the circuit.

Practical implementations of \acp{dd} rely on a synergetic pair of data structures: a \ut for node persistence and a \ct for operation tracking. Such structures are highly effective for the recursive processing characteristic of tree-based \acp{dd}, where operations are decomposed into subtree evaluations~\cite{HZL22}. When an operation involving specific subtrees has been previously encountered, the result is extracted from the \ct rather than recomputed. This approach is especially beneficial when dealing with large-scale sub-structures, as it yields a drastic reduction in the total number of required operations.

To bridge the gap between the topological flexibility of \acp{tn} and the symbolic efficiency of \acp{dd}, \acp{tdd} were recently introduced~\cite{HZL22}. This hybrid approach allows for the representation of complex tensor indices within a diagrammatic structure, facilitating more sophisticated optimization strategies. Recent studies have demonstrated that the performance of such hybrid simulators is highly sensitive to the global indexing strategy~\cite{ZSK24}, a challenge that the present work addresses through novel ordering heuristics and low-level architectural refinements. Nevertheless, the viability of this approach depends on several essential operations that bridge the gap between \acp{dd} and \acp{tn}. Of these, tensor contraction is perhaps the most fundamental; its specific implementation details and optimizations are further explored in~\cite{HZL22}.

\begin{figure}[ht]
    \centering
    \includegraphics[width=\linewidth]{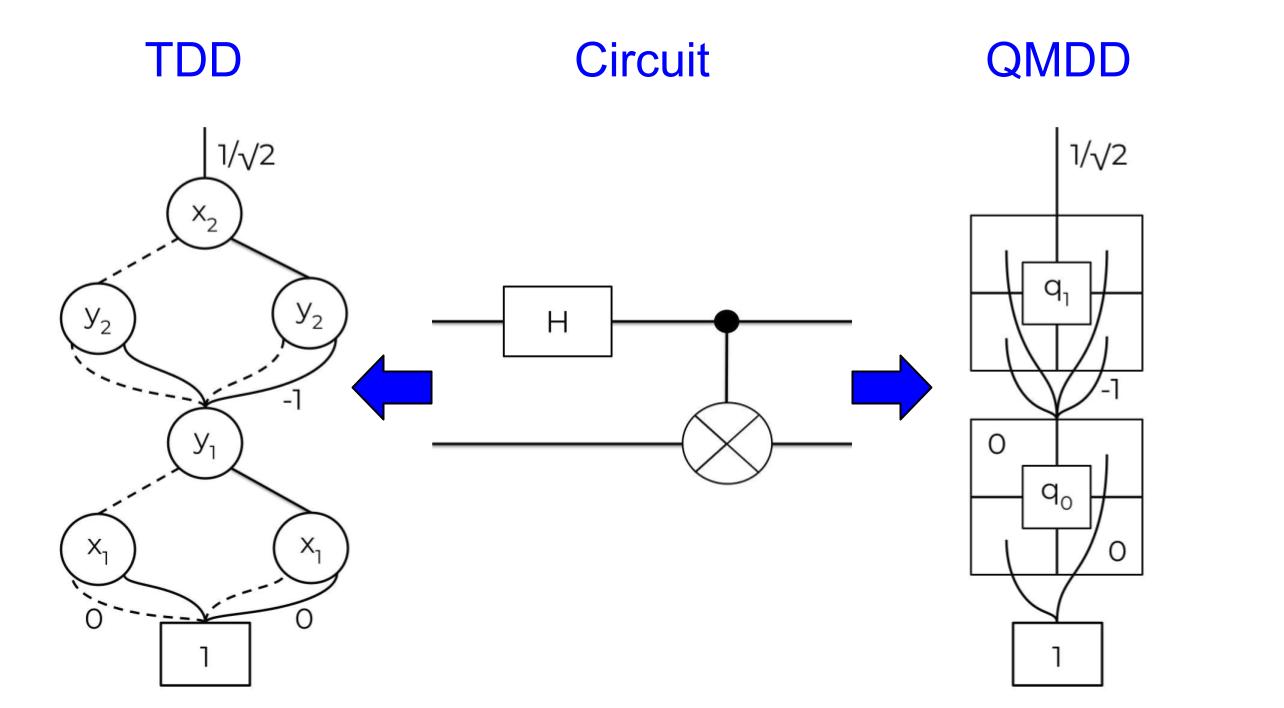}
    \caption{Representation of the matrix that represent the functionality of the full circuit as \ac{tdd} on the left and as \ac{qmdd} on the right}
    \label{fig:dds}
\end{figure}

Figure~\ref{fig:dds} shows two different DD representations of the same circuit. \acp{tdd} encapsulate \ac{tn} indices through a nodal hierarchy where each element branches into 2 successors, corresponding to the node's binary valuations. The relationship between a node and its successors is defined by weighted edges; these weights act as scalar multipliers for the underlying sub-structures. In practice, any edge lacking an explicit numerical weight is interpreted as having a weight of 1, thereby streamlining the visual representation of the network. \ac{qmdd} work in a similar fashion, with the difference that each node have 4 successors instead of 2.

While the original \acp{tdd} described in~\cite{HZL22} were developed in Python, this work utilizes \ac{ftdd}, an enhanced implementation written in C++~\cite{ZSK24}. This high-performance iteration introduces several architectural refinements, including edge-centric data structures and the integration of classic \ac{bdd} optimization techniques~\cite{BRB91}. Furthermore, by employing a near-optimal contraction ordering via the \cotengra library, \ac{ftdd} achieves acceleration factors of up to 175x compared to its Python-based predecessor.

\subsection{Hash Tables}

Hash tables are designed to facilitate near-instantaneous data access, providing an average $O(1)$ complexity for lookup, insertion, and deletion. Initially conceptualized in~\cite{Luh53} as a method to accelerate data retrieval through numerical key transformation, the structure was later formalized in~\cite{Don99}, establishing the design heuristics for modern hashing. In contemporary computing, hash tables are indispensable, particularly within \ac{dd}-based frameworks used for symbolic simulation. Their performance is fundamentally tied to the quality of the hash function (which distributes keys across an underlying bucket array) and the efficiency of its collision-handling mechanism. This dual-component design offers a clear performance advantage over logarithmic search structures such as balanced trees.
Building upon these principles, \cite{BRB91} proposed a robust architecture that remains the gold standard for \ac{bdd} packages. This framework relies on two primary hash-table-based structures, each serving a unique role in the system's operational logic:

Most contemporary \ac{bdd} packages are built upon the architectural foundations laid in~\cite{BRB91}, which rely on two specialized hash tables to optimize system performance:
\begin{enumerate}
    \item \ut: It guarantees the canonical representation of the \ac{dd} by managing node uniqueness. It is typically indexed by a tuple representing the node’s properties; if a node with an identical structure already exists, the table returns a reference to the existing instance. This approach prevents structural isomorphism and ensures that the total memory footprint remains as small as possible.

    \item \ct: This structure serves as a dedicated cache to store the outcomes of previously executed operations. By indexing entries as (operation, operands, result), the \ct allows the system to bypass redundant computational paths. This feature is indispensable for maintaining efficiency during the simulation of complex quantum circuits, where repetitive, and probably costly, sub-operations are frequent.

\end{enumerate}

\begin{figure}[ht]
    \centering
    \includegraphics[width=\linewidth]{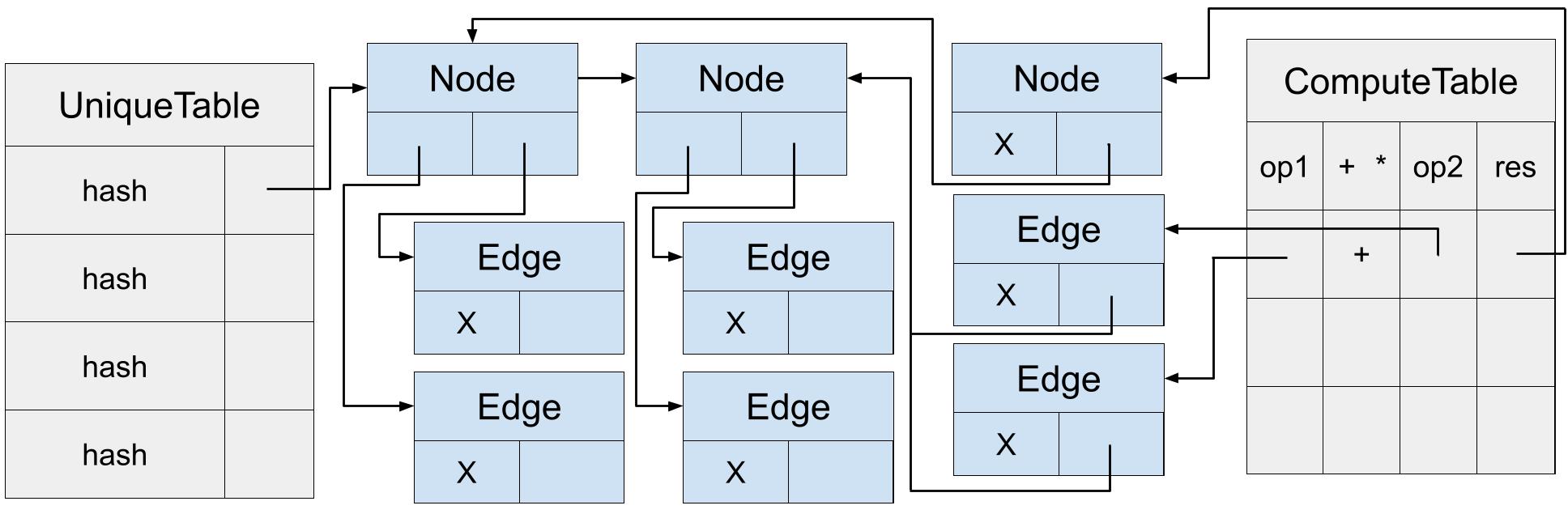}
    \caption{Example of how the \ut and \ct works. The X represents an arbitrary complex number and the arrows represents memory pointers.}
    \label{fig:hst}
\end{figure}

Figure~\ref{fig:hst} shows an example of how this structures are implemented in the \ac{ftdd} tool. They use 2 structures to represent the \ac{tdd}: The \texttt{Node} class, which represent the circle of the Figure~\ref{fig:dds} and the \texttt{Edge} class, which represents the edges. The edge class is used to store the common factor, further improving the detection of nodes structurally identical. In this implementation, the \ut stores in each entry a linked list of the unique nodes that have the same hash value. On the other hand, in the implementation of the \ct only one value per entry can be stored. If a collision takes place, then the last result overwrites the old one. It is also important to notice that, while the \ut stores pointers to nodes, the \ct stores 2 pointers to edges (that represents the operands) and a pointer to a node (that represents the result of the operation).

Given their lineage, \acp{ftdd} inherit these implementation strategies, adopting the same fundamental hash table mechanisms described in~\cite{BRB91} to ensure both canonicity and operational speed.

\subsection{Order of Global Indices}


The computational efficiency of a \ac{dd} is fundamentally contingent upon the sequence in which its node level are processed. This strongly depends on the kind og \ac{dd} that is being used. For example, within the state-of-art \ac{qmdd}, each diagram level typically corresponds to a physical qubit, often employing an interleaved row and column variable structure to represent unitary transformations~\cite{HZL22}. Conversely,  within the framework \ac{tdd}, this ordering (the spatial arrangement of global tensor indices) represents a critical determinant of both execution performance and memory usage.
The levels in a \ac{tdd} represent abstract tensor indices, effectively decoupling internal diagrammatic levels from physical qubit labels. This architectural shift allows \acp{tdd} to represent tensors of arbitrary order while remaining unaware of the underlying physical qubit topology. This decoupling introduces a dual-natured paradigm for quantum circuit simulation. On the one hand, it provides a significantly more flexible search space for symbolic representation, potentially yielding more compact structures than those achievable through qubit-centric diagrams. On the other hand, such flexibility increases the complexity of identifying an optimal ordering, as the relationship between the logical quantum circuit and its abstract tensor-level representation is non-trivial and non-linear~\cite{HBS22}.

\begin{figure}[ht]
    \centering
    \includegraphics[width=\linewidth]{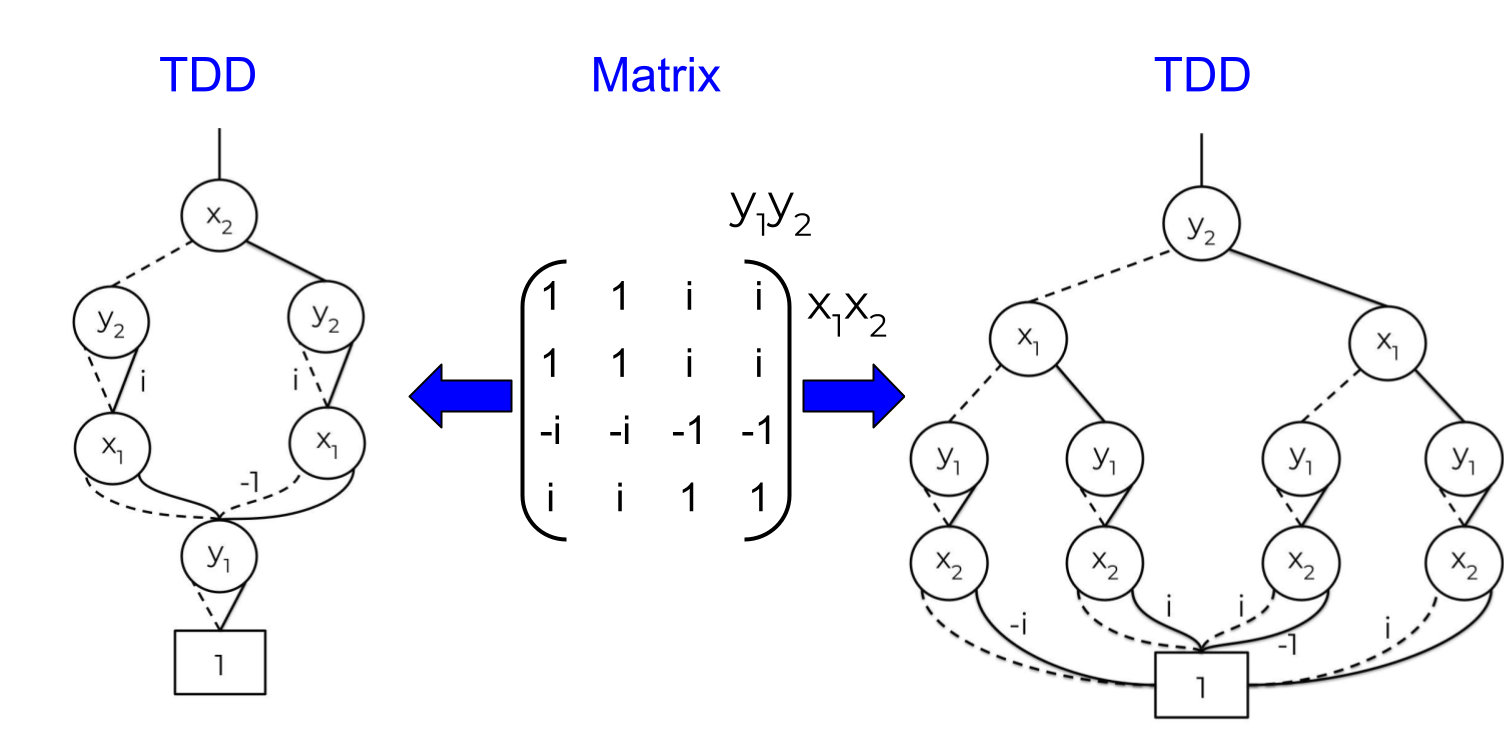}
    \caption{Difference in the \ac{tdd} representation of the matrix with two different index ordering. The left one uses the index order $x_2 > y_2 > x_1 > y_1$ and the right one uses $y_2 > x_1 > y_1 > x_2$.}
    \label{fig:ord}
\end{figure}

The significance of the ordering problem is rooted in the mathematical property of canonicity. For a fixed linear ordering of indices $\pi(\mathcal{I})$, any given tensor possesses a unique, deterministic, and reduced \ac{tdd} representation. Consequently, the size of the diagram (defined by its amount of nodes) is a direct function of the chosen index permutation. Figure~\ref{fig:ord} shows the impact of a good index ordering compared to a bad one. The left-hand \ac{tdd} has a total of 6 nodes, while the right-hand one has 11. That means, in this example, we almost halved the number of nodes only by changing the index order of the \ac{tdd}. The optimization challenge can thus be formally defined: given a set of indices $\mathcal{I}$ associated with a \ac{tn}, the objective is to identify a permutation $\pi(\mathcal{I})$ that minimizes the total number of nodes in the resulting TDD. Because the number of possible permutations scales factorially ($n!$) with the number of indices, an exhaustive search for the absolute optimum is computationally intractable for large-scale systems. This challenge is situated within the NP-hard complexity class, mirroring the well-documented variable ordering problem in classical Ordered Binary Decision Diagrams (OBDDs)~\cite{HBS22}.

Methodologies to address this optimization bottleneck are generally categorized into two primary families:
\begin{itemize}
    \item \textbf{Static Ordering:} Determine the index sequence prior to the commencement of the contraction or simulation process. Static approaches typically rely on the initial topology of the \ac{tn} or user-defined naming conventions. Notably, the seminal \ac{tdd} implementation by Hong et al. \cite{HZL22} utilizes a baseline static alphanumeric ordering as its default configuration. While computationally inexpensive, static methods often fail to capture the dynamic correlations and entanglement structures that emerge during successive tensor contractions.
    \item \textbf{Dynamic Reordering:} These techniques involve the real-time adjustment of the index sequence during simulation to mitigate memory growth. Derived from the BDD literature, dynamic strategies such as Rudell's sifting algorithm~\cite{Rud93} aim to reduce diagram size via local swaps of adjacent levels. Although sifting is effective in managing ``memory explosion" in complex algorithms, it incurs substantial runtime overhead and may face numerical stability challenges due to the accumulation of floating-point errors in complex edge weights.
\end{itemize}
The \ac{ftdd} framework uses one particular static ordering based on the alphanumeric ordering of the indices. The authors of the framework used this particular order to demonstrate the advantage of adapted index order to the needs of a circuit in contrast to the fixed index order in the \acp{qmdd} tools. They demonstrate that this particular order is better with QFT~\cite{HZL22}.

\subsection{Simulation Methodologies}



Classical quantum circuit simulation methods can be broadly classified into exact and approximate approaches. Exact methods preserve the complete mathematical representation of the quantum state throughout the simulation, apart from the unavoidable effects of finite-precision arithmetic. Approximate methods, in contrast, reduce the computational or memory requirements by discarding part of the represented information or by estimating only selected properties of the quantum system. This distinction is particularly relevant to the present work, since \ac{ftdd} performs exact simulation and seeks to improve scalability through structural compression, memory management, and index ordering rather than through numerical approximation.

\subsubsection{Exact Simulations}


State-vector simulation is the most direct exact approach. An \(n\)-qubit quantum state is explicitly represented by an array containing \(2^n\) complex amplitudes, and each quantum gate is simulated through the corresponding linear-algebra operation. This representation is general and provides direct access to the complete quantum state, but its memory requirements grow exponentially with the number of qubits~\cite{NiC11}. The reason for that is based in the dimensionality of the Hilbert space associated with an N-qubit system, whose statevector requires the specification of $2^n$ complex amplitudes. As $n$ scales, the explicit storage and manipulation of this vector become prohibitive due to the exponential growth of memory requirements \cite{Fey82}, as it will require to explicitly storing an array of $2^n$ complex numbers in RAM \cite{RMR07}. To illustrate the problem, given a system of just 50 qubits in double precision, to store the value of the statevector it will requires:
\begin{equation}
    2^{50} \times 8 \times 2 = 16 \text{ PiB}
\end{equation}
Consequently, exact state-vector simulation is generally restricted to approximately 40-50 qubits, even on large high-performance computing infrastructures~\cite{de2025universal, PGN17}.

\acp{dd} provide an alternative exact representation. Rather than storing all amplitudes explicitly, they encode the quantum state or operator as a directed acyclic graph and share equivalent substructures. For circuits with high regularity, symmetry, or repeated computational patterns, this structural compression can substantially reduce the size of the representation and allow the exact simulation of systems beyond the practical limits of conventional state-vector methods. However, this advantage depends on the amount of redundancy that can be exploited. For irregular, highly entangled, or pseudo-random circuits, the number of nodes may still grow exponentially, increasing both memory consumption and execution time.

\ac{tn} methods can also perform exact simulation when all tensor operations are carried out without truncation and all resulting contributions are retained. In this case, the main limitation is the size of the intermediate tensors generated during contraction, which depends strongly on the network structure and on the selected contraction path. \ac{ftdd} belongs to this exact simulation category: it combines \ac{tn} contraction with the structural compression provided by \acp{dd}, without discarding tensor components or approximating the final result.

\subsubsection{Approximate Simulation}

Approximate methods trade part of the accuracy or completeness of the simulation for improved scalability. \acp{tn} are frequently used in this regime because their internal dimensions can be explicitly controlled, thereby limiting the growth of intermediate representations. One common strategy is bond-dimension truncation. During the evolution of the circuit, singular-value decomposition can be used to identify and discard components with a small contribution to the represented state. This reduces memory usage and execution time, although it introduces a loss of fidelity whose magnitude depends on the selected truncation threshold.

Slicing constitutes another strategy for reducing peak memory consumption~\cite{huang2020classical}. Selected tensor indices are fixed, producing a collection of smaller and independent contraction problems. Slicing is mathematically exact when all slices are computed and their results are correctly aggregated. However, in large-scale simulations, the number of slices may itself become prohibitively large. Practical implementations may therefore evaluate only a subset of the slices, sample them stochastically, or discard contributions considered negligible, in which case the resulting simulation becomes approximate~\cite{chen2018classical, schuch2007computational, gray2021hyperoptimized}. A further alternative is to estimate amplitudes, probabilities, or expectation values through sampling techniques, including Monte Carlo~\cite{MaS08} methods over selected computational paths. These approaches avoid constructing the complete state representation, but their results are statistical estimates whose accuracy
depends on the number of samples. Other option is to use Singular Value Decomposition (SVD), \acp{tn} compute the entanglement spectrum and actively discard the smallest singular values below a threshold \cite{schollwock2011density}. This artificially fixes the bond dimension, reducing memory but degrading the fidelity of the quantum state.

Unlike these approximate approaches, \ac{ftdd} does not rely on truncation, partial slicing, or statistical sampling. Its scalability therefore depends on the compactness of the \ac{tdd} representation, the contraction strategy, the global ordering of tensor indices, and the efficient management of the internal data structures created during the simulation.

\section{FTDD Memory Optimization}
\label{sec:memory}

Based on the foundational evaluation of \acp{ftdd} in a diverse range of quantum circuits and contraction strategies~\cite{LBC25}, the original implementation exhibited several functional and memory-management limitations that reduced its robustness and scalability during quantum circuit simulation. Before addressing the main memory bottlenecks of the framework, a set of preliminary improvements was introduced to consolidate the implementation and provide a reliable basis for the subsequent optimizations.

\subsection{Preliminary Framework Improvements}

Previous versions of \ac{ftdd} presented inconsistencies in the detection and propagation of open tensor indices. During successive \ac{tn} transformations, these inconsistencies could produce malformed intermediate representations and undermine the reliability of the resulting \ac{tn}. The revised implementation therefore ensures that open indices are consistently identified, validated, and updated throughout the simulation pipeline.

In addition, a standardized high-level simulation interface was developed to simplify the execution of quantum circuit simulations within \ac{ftdd}. This interface encapsulates tool initialization, circuit preprocessing, and \ac{tn} construction within a unified execution procedure. By reducing the amount of user-side configuration and enforcing a consistent sequence of operations, the interface simplifies the use of \ac{ftdd} and improves methodological reproducibility. These preliminary improvements provide the implementation basis for the memory-management and index-ordering optimizations introduced in the following sections.

\subsection{Memory-Management Redesign}

The scalability of \ac{ftdd} is strongly influenced by the large number of internal elements generated during recursive \ac{tdd} operations and tensor contractions. In the original implementation, memory allocation followed a dynamic and unbounded growth model. At the same time, the lifecycle of nodes was not tracked with sufficient precision, and inactive nodes were not efficiently reused. As a result, complex contractions could lead to excessive memory consumption, repeated memory allocation, and inconsistencies between the internal structures used by the simulator.

To address these limitations, the memory-management mechanisms of \ac{ftdd} were revised through explicit node reference tracking, safe reuse of inactive nodes, selective operation caching, and a hardware-aware sizing policy for the main internal tables. These mechanisms seek to control memory growth while balancing lookup efficiency, garbage-collection overhead, and execution time.

In \ac{ftdd}, the garbage-collection mechanism scans the internal tables to identify nodes that are no longer in use and can therefore be reclaimed or made available for reuse. Its cost depends on the size and organization of the tables that must be traversed. Consequently, table sizing, node tracking, and node reuse are closely related aspects of the memory-management model.

\subsection{Node Lifecycle and Internal Data Structures}

equivalent nodes are represented by a single canonical instance, whereas the \ct stores the results of previously executed operations to avoid redundant computation. 

In the original implementation, the interaction between these two structures could produce inconsistent node lifecycles. A node removed from the \ut could still be referenced from the \ct, creating dangling references and potentially leading to invalid pointer accesses or inconsistencies during search operations. To prevent this situation, the revised \ct introduces explicit reference tracking. Nodes involved in cached operations cannot be reclaimed while they remain referenced by the table. This mechanism preserves pointer integrity and maintains consistency between cached operations and the node structures on which they depend.

The \ut was also modified to support the safe reuse of nodes that are no longer semantically active. Previous versions relied on a one-use lifecycle that increased memory pressure during large contractions. The new recycling mechanism allows inactive nodes identified by the garbage collector to be reused, reducing the frequency of new memory allocations and improving the memory footprint of the simulator. 

\subsection{Selective Operation Caching}

The criteria used to store operations in the \ct were also revised. In the new implementation, an operation is stored in or retrieved from the \ct only when the combined depth of the participating decision diagrams exceeds a predefined threshold. In the current implementation, this threshold is set to 3. This policy reserves table capacity for computationally more expensive operations and avoids using cache space for operations whose reuse is expected to provide a limited benefit.

The physical memory layout of the entries was also optimized so that the indices involved in an operation are stored close to their corresponding buckets, reducing cache misses during frequent table accesses. This caching policy also affects memory management because cached operations retain references to their associated nodes. Therefore, storing operations that are not subsequently reused can increase memory consumption and reduce the effectiveness of node reclamation.

\subsection{Hardware-Aware Table Sizing}

A further limitation of the original implementation was the sizing policy used for the \ut and \ct structures. In particular, the memory allocated for the \ut scaled exponentially with the number of qubits, leading to unpredictable resource consumption even for moderate circuit sizes.

The revised implementation introduces a hardware-aware sizing policy that determines the dimensions of both tables according to the available physical memory. \ac{ftdd} employs a 32-bit FNV hashing scheme~\cite{HaM24}, and table sizes of the form \(2^x - 1\) are used. A hardware-dependent parameter \(N\) determines the number of buckets assigned to each structure. The \ut is initialized with \(2^N - 1\) buckets, whereas the \ct{} is initialized with \(2^{N-1} - 1\) buckets. The value of \(N\) is selected so that the combined memory footprint of both tables occupies approximately \(1\%\) of the available system memory. This resource-aware allocation strategy allows \ac{ftdd} to adapt its internal capacity to the host system and prevents the table allocation itself from producing uncontrolled memory growth.

The selected table sizes also influence garbage-collection cost. Larger tables may reduce bucket density and improve lookup performance, but they require a larger memory space to be traversed when inactive nodes are identified. Smaller tables reduce the number of buckets, but may increase collisions and the number of entries associated with each bucket. The sizing of the internal tables introduces a trade-off between memory consumption and execution efficiency. Larger tables provide more capacity and may reduce hash collisions, but they also increase the memory footprint and the cost of traversing the tables during garbage collection. More restrictive table sizes reduce the allocated memory, but may decrease lookup efficiency and the effectiveness of the internal caching mechanisms.

The \ct introduces an additional trade-off. Increasing its capacity may improve the reuse of intermediate operations, but the cached results also retain the nodes associated with those operations. If these operations are not frequently reused, the additional memory consumption may not be compensated by a reduction in execution time. For this reason, the revised implementation combines hardware-aware table sizing, selective caching, explicit reference tracking, garbage collection, and node recycling. The experimental evaluation presented later compares different table-sizing policies and analyzes their impact on memory consumption, garbage-collection efficiency, and execution time across quantum circuits with different structures and topologies.
\section{Index Ordering Strategies}
\label{sec:order}
As said before, the efficiency of \ac{dd}-based simulation is heavily predicated on the linear arrangement of tensor indices. To transcend the limitations of basic \alp ordering, we explore other two heuristics with different features: one rooted in the topological properties of the \ac{tn} and another derived from its operational dynamics. In this sections we will revise the three ordering methods.

\subsection{Alphanumeric}
The original implementation of \ac{ftdd} uses a single static index-ordering method, which we refer to as the \alp ordering. After converting the quantum circuit into a \ac{tn}, \ac{ftdd} assigns a name to each tensor index and orders the indices according to these names. Therefore, the resulting order depends directly on the naming convention used during the circuit-to-\ac{tn} conversion.

\begin{figure}[ht]
    \centering
    \includegraphics[width=0.9\linewidth]{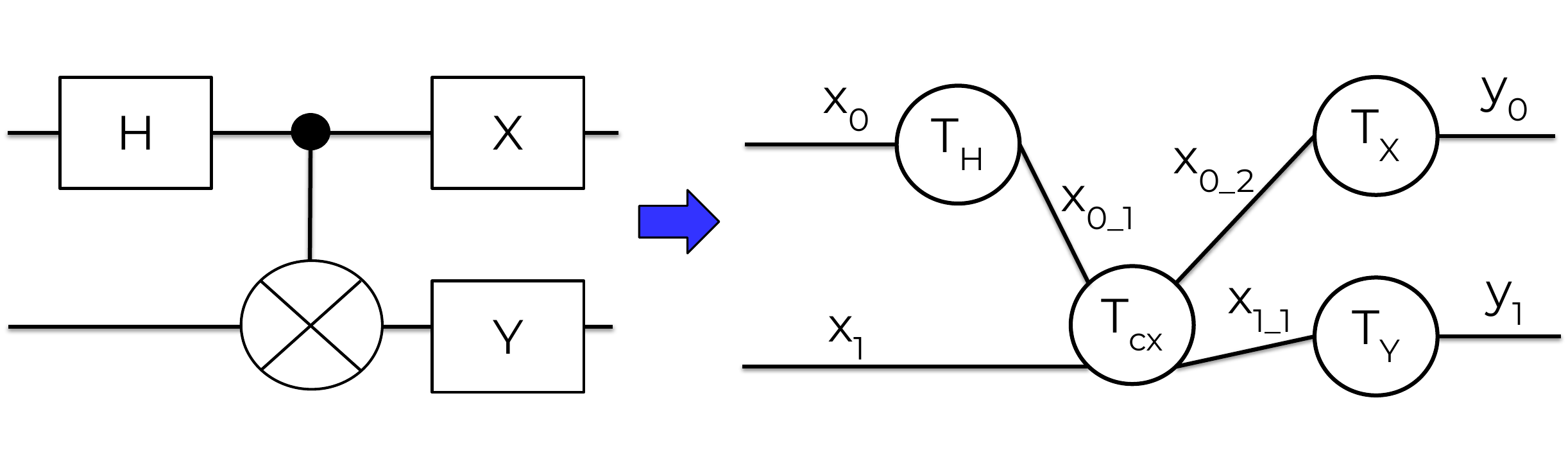}
    \caption{Conversion of a two-qubit circuit into a \ac{tn} and the index
    names assigned by \ac{ftdd}. Each gate is represented by a tensor, and
    each segment of a qubit line corresponds to a tensor index.}
    \label{fig:alp}
\end{figure}

Figure~\ref{fig:alp} illustrates this naming convention. The quantum circuit is shown on the left, while its corresponding \ac{tn} representation is shown on the right. Each quantum gate is represented by a tensor, and each segment of a
qubit line becomes an index connecting two tensors or an open index of the network.

For qubit line $n$, the input index is denoted by $x_n$ and the output index by $y_n$. Intermediate indices are denoted by $x_{n\_k}$, where $k$ indicates the position of the index along qubit line $n$ as the circuit is processed from
left to right.


Using the names assigned in this example, the \alp index order is
\[
x_0 > x_1 > x_{0\_1} > x_{0\_2} > x_{1\_1} > y_0 > y_1.
\]

\subsection{Reverse Cuthill-McKee (RCM)}
The Reverse Cuthill-McKee (RCM) algorithm is a classical graph-theoretic heuristic originally conceived for bandwidth reduction in sparse symmetric matrices~\cite{CuM69}. In the context of \acp{tdd}, RCM aims to localize dependencies by minimizing the distance between related indices (that means, indices that in any given moment are used at the same time in any \ac{dd} of the contraction process) within the diagram's hierarchy. The traditional application of RCM to \ac{tn} necessitates the construction of an interaction graph $G = (V, E)$, where the vertex set $V$ represents all indices of the network. An edge $e = (i, j)$ exists in $E$ if, and only if, indices $i$ and $j$ coexist within at least one tensor. The algorithm proceeds as follows:
\begin{enumerate}
    \item \textbf{Pseudo-peripheral Node Identification:} The process begins by selecting a starting node with a minimal degree, typically representing a boundary index in the network topology.
    \item \textbf{Level Structure Construction:} A Breadth-First Search (BFS) is initiated. During the traversal, neighbours of the current node are visited in increasing order of their degrees. This ensures that the most ``constrained" parts of the graph are processed in a tight sequence.
    \item \textbf{Reversal:} The resulting sequence is reversed to yield the final RCM order.
\end{enumerate}
By reducing the graph's bandwidth, RCM ensures that indices sharing a common tensor are positioned in proximal levels of the \ac{tdd}. This spatial proximity is instrumental in fostering structural sharing within the \ut and thus suppressing the exponential growth of nodes in intermediate stages of the representation. Specifically, structural sharing refers to the canonical property of \acp{dd} where isomorphic subgraphs are uniquely instantiated in memory. By grouping interdependent indices tightly within the hierarchy via RCM, the probability of encountering identical sub-functions in the tensor contractions increases. Consequently, instead of allocating redundant nodes, multiple parent nodes across different paths redirect their pointers to a single, globally shared sub-diagram in the \ut, drastically reducing the memory footprint.

\subsection{Path-based Ordering}

In our new proposed heuristic for ordering the indices, we minimize the distance in the index order of the indices involved in each contraction during the simulation process. Path-based Ordering is a static strategy that leverages the specific sequence of operations planned for the simulation (contraction path), as opposed to RCM, that is an agnostic, static heuristic based solely on the initial network topology.  Theoretically, this algorithm can be understood as an instance of \textit{cut-width} minimization for the index interaction graph, guided by the contraction path. As demonstrated in~\cite{MaS08}, the complexity of contracting a \ac{tn} is linked to the \textit{treewidth} of the network's line graph. However, the size of a \ac{dd} depends on a different parameter: the linear ordering of its variables. While the contraction order optimizes for \textit{treewidth}, \textit{Path-based Ordering} operates on a complementary layer, deriving a linear ordering that seeks to minimize the \textit{pathwidth} or \textit{cut-width} induced by said plan.

This heuristic relies on the life cycle of an index. Each index in a network is either a \textit{contracted index} (internal) or a \textit{free index} (external). The importance of an index to the \ac{tdd} size is highly correlated with when it is eliminated from the system. The algorithm facilitates this through the following logic:

\begin{itemize}
    \item \textbf{Contraction Scheduling:} Given a contraction path, the algorithm simulates the contraction steps to identify the precise moment (step $s$) each index $i$ is consumed.
    \item \textbf{Priority Mapping:} Indices are ordered according to their ``elimination time". Indices that are contracted early in the simulation are prioritized for the upper levels of the \ac{tdd}. This allows the simulator to resolve and ``prune" the corresponding branches of the decision diagram as soon as the algebraic operation is completed.
    \item \textbf{Persistence Handling:} A secondary sorting criterion is applied based on the ``last seen" step of an index, ensuring that indices which persist longer in the working set of tensors are pushed toward the lower levels of the diagram.
\end{itemize}

This greedy approach is highly effective because it seeks to locally minimize the number of active nodes at each level of the diagram. Recently, work on FeynmanDD~\cite{WCY25} has formalized this relationship, proving that the size of a \ac{dd} is exponential with respect to the \textit{linear rank-width} (a variant of \textit{cut-width}) of the circuit graph. By aligning the \ac{tdd} hierarchy with the temporal evolution of the simulation, the diagram acts as a compact state space that prevents the retention of obsolete information, achieving speedups of up to 13x compared to traditional static orderings. 

As detailed in Algorithm~\ref{alg:path}, the Path-based ordering method simulates the entire contraction sequence $P$ to map the temporal footprint of each index. We track two critical variables for every index: $FirstStep$, which marks the earliest point at which an index is eliminated via contraction, and $LastSeen$, which identifies the final operation where an index remains active in the intermediate tensor set. The indices are then sorted primarily by their $FirstStep$. This ensures that indices consumed early in the simulation occupy the uppermost levels of the \ac{tdd}. By resolving these variables first, the diagram can effectively prune redundant sub-graphs early in the recursive process. For indices that are contracted in the same step, the $LastSeen$ value serves as a secondary tie-breaker to further refine the hierarchy based on index persistence.

\begin{algorithm}[htp]
\SetAlgoLined
\DontPrintSemicolon
\KwIn{ \\ \hspace{1cm} $\mathcal{T} = \{T_1, T_2, \dots, T_m\}$ set of tensors \\
\hspace{1cm} $P = \{(i_1, j_1), \dots, (i_{m-1}, j_{m-1})\}$ contraction path }
\KwOut{ \\ \hspace{1cm} $\mathcal{O}$ ordered list of indices }

\BlankLine
\tcp{Get all indices of the tensor network}
$WorkingSet \gets \{Indices(T) \mid T \in \mathcal{T}\}$ \;
\tcp{Step where an index is first contracted}
$FirstStep \gets \emptyset$

\tcp{Last step where an index is involved}
$LastSeen \gets \emptyset$ 

\BlankLine
\tcp{For every contraction pair}
\For{$s \gets 1$ \KwTo $|P|$}{
    \tcp{Get the position of the tensors involved}
    $(a, b) \gets P[s]$ \;
    \tcp{Indices for which the contraction was done}
    $Contracted \gets WorkingSet[a] \cap WorkingSet[b]$ \;
    \tcp{Indices of the resulting tensor}
    $Surviving \gets (WorkingSet[a] \cup WorkingSet[b]) \setminus Contracted$ \;
    \tcp{For every contracted index}
    \ForEach{$idx \in Contracted$}{
        \tcp{If not used before, then marked as first use}
        \If{$idx \notin FirstStep$}{
            $FirstStep[idx] \gets s$ \;
        }
        \tcp{Mark every contracted index as last seen}
        $LastSeen[idx] \gets s$ \;
    }
    \tcp{Mark every resulting index as last seen}
    \ForEach{$idx \in Surviving$}{
        $LastSeen[idx] \gets s$ \;
    }
    
    Update $WorkingSet$ by replacing $WorkingSet[a]$ and $WorkingSet[b]$ with $Surviving$ \;
}

\BlankLine
\tcp{Get all indices of the tensor network}
$AllIndices \gets \bigcup_{T \in \mathcal{T}} Indices(T)$ \;
\tcp{Calculate the free indices}
$FreeIndices \gets AllIndices \setminus \text{domain}(FirstStep)$ \;
\tcp{Order de non-free indices first by first usage and the when last seen}
$\mathcal{O}_{contracted} \gets \text{sort indices } idx \in \text{domain}(FirstStep) \text{ by } (FirstStep[idx], LastSeen[idx])$ \;
$\mathcal{O}_{free} \gets \text{Rearrange } FreeIndices \text{ according to adjacency}$ \;

\Return $\mathcal{O}_{contracted} + \mathcal{O}_{free}$
\caption{Path-based Index Ordering}
\label{alg:path}
\end{algorithm}
\section{Experimental Evaluation of Memory and Indexing Optimizations}
\label{sec:res}
This section introduces the experimental methodology, followed by a comprehensive analysis of the resulting performance data. Our primary objective is two-fold: to evaluate the spatial and temporal efficiency of our enhanced \ac{ftdd} implementation relative to its predecessor and to evaluate the effect of different index ordering methods. 
The aim is to demonstrate the scalability gains afforded by our hardware-aware memory management and optimized index ordering strategies.

\subsection{Experimental Environment and Methodology}
The experimental evaluation was conducted on a high-performance computing node equipped with two Intel Xeon 6418H processors (24 cores each), 2 TB of RAM, and four NVIDIA H100 (80 GB) GPUs, running under Ubuntu Linux (kernel 5.4.0-72-generic). Despite the node’s powerful parallel computing architecture, it was selected primarily for its large main memory, which represents the primary limiting factor in the simulations. Owing to the sequential nature of the simulator, all experiments were executed on a single CPU core. The benchmarking process was divided into two primary phases: an initial assessment of the architectural refinements in the latest \ac{ftdd} version (Section~\ref{subsec:rimp} and~\ref{subsec:rtbl}), followed by a comprehensive comparative analysis of the different index orders (Sections~\ref{subsec:order}). 

The benchmark suite was sourced from MQTBench~\cite{QBW22}, from which nine representative circuits were selected based on the topological classification metrics proposed in~\cite{VRV26}. These circuits cover a broad spectrum of structural characteristics, exhibiting significant variations in depth, gate count, entanglement, and computational complexity

Standardized experimental conditions were maintained across all trials. Simulations focused on state-vector evolution, utilizing a fixed input state $\ket{0}$ while maintaining open output indices. To evaluate performance across different optimization regimes, two distinct contraction strategies were employed: the \seq method, selected as a baseline for its simplicity, and the \cotengra framework, representing a state-of-the-art optimization paradigm. For the latter, \cotengra was allocated one hour of wall-clock time to identify a contraction path using the \textit{combo} minimization parameter with 56 concurrent threads. No circuit-specific preprocessing was applied prior to simulation. Each simulation trial was subject to a strict two-hours time constraint. Exceeding the time limit is marked with $T.O$, while exceeding the memory available is marked with $O.M$. In addition, each circuit is limited by their number of qubits available in the benchmark suite (marked with an $*$), or the limit of 100 qubits imposed by other internal data structures of the \ac{ftdd} tool (marked with a \dag). Experiments proceeded iteratively, starting from the minimum qubit count and scaling upward until the simulator reached either a temporal timeout or memory exhaustion. This approach allows for a precise determination of the scalability limits of the \ac{ftdd} tool under hardware-constrained conditions.

In order to evaluate the performance of the tool, several metrics were used to better understand its behaviour. The metrics used are the following:
\begin{enumerate}
    \item \textbf{Time (s)}: Execution time measured in seconds used only to contract the \ac{tn} of the circuit. Other processes such as loading the circuit are not measured as it does not depends on our tool.
    \item \textbf{MeM (GiB)}: Real usage of the RAM memory used by our tool. As nodes are never erased, the amount of memory used at the end of the contraction process is also the maximum amount of memory used along the simulation.
    \item \textbf{Max\_bck}: Maximum number of nodes that are in the same entry of the \ut.
    \item \textbf{hit ratio}: The percentage of successful operation lookups in the \ct structure.
    \item \textbf{node counts}: Total active number of nodes that are used during the contraction process.
\end{enumerate}

\subsection{Evaluation of Memory Optimization Strategies}
\label{subsec:rimp}

To validate the efficacy of the architectural and memory-management enhancements described in Section~\ref{sec:memory}, we conducted a comparative analysis between the original \ac{ftdd} implementation and our optimized framework. Our evaluation focused on two primary metrics: memory scalability and execution time, both assessed across a representative suite of quantum circuits ranging from small-scale structures to high-depth, high-entanglement instances.



Table~\ref{tab:comp} presents the empirical results of this evaluation across 9 quantum circuit topologies, scaling the number of qubits ($n$). In the table, only the \seq contraction plan is analysed, but the conclusions are equivalent if \cotengra is used. Bold letter is used for the best case among both versions of the framework for each of the three metrics included. At a macroscopic level, Table~\ref{tab:comp} reveals distinct behaviour between both versions. In terms of temporal performance, the original implementation exhibits a sharp exponential growth across most benchmarks, rapidly approaching or exceeding the one-hour execution threshold (e.g., in QPE and QWalk), or collapsing due to memory exhaustion before completing the time limit. In contrast, the optimized framework stabilizes temporal growth completely in highly regular circuits, allowing execution up to 100 qubits with QFT circuits. 

Regarding memory utilization, the baseline \ac{ftdd} framework demonstrates a steep, unconstrained growing slope. Our hardware-aware, fixed-footprint model significantly controls its memory usage. A notable exception to this general trend occurs in the QPE and QWalk benchmarks, where the memory consumption profiles and growth slopes remain closely matched between both versions (e.g., at $n=30$ for QPE, the original requires 115.21~Gb while the improved version utilizes 111.77~Gb).

By analyzing these macro-trends through the topological features of the circuits, the benchmarks can be categorized into three distinct behavioral profiles governed by their underlying quantum structures, with corresponding circuit types grouped adjacently in Table~\ref{tab:comp}:
\begin{itemize}
    \item \textbf{Highly Structured/Regular Topologies (QFT, GHZ, Graph):} These circuits possess repetitive structural patterns and localized operations. In this regime, our optimized framework achieves an ideal stabilization; memory and execution time flatline almost completely (e.g., QFT scales from 30 to 100 qubits consuming a constant $\approx 13.12$~Gb in roughly 25 seconds), while $Max\_bck$ drops to 1, proving optimal hash distribution.
    \item \textbf{Algorithmic/Balanced Topologies (QPE, QWalk):} These intermediate-complexity circuits feature medium repetitions of their structural patterns and localized operations. Here, both versions exhibit similar memory growth, but the improved framework maintains slightly better or equal execution times while preventing unpredictable allocation spikes.
    \item \textbf{High-Depth/High-Entanglement Topologies (RA, AE, QNN, RQC)}: These complex, pseudo-random, or variational networks represent the most demanding regimes. In these circuits, a clear trade-off emerges. The original version suffers from catastrophic memory inflation (e.g., QNN at 25 qubits hits 309.13~Gb and RQC at 18 qubits requires 104.17~Gb). Our framework drastically reduces this footprint (bringing them down to 67.45~Gb and 3.40~Gb, respectively), preventing $O.M.$ failures. However, this strict memory containment comes at the expense of a significant temporal penalty and a sharp inflation in $Max\_bck$ (e.g., spiking to 210 in RQC at 12 qubits), caused by the heavy computational overhead of explicit reference tracking and node lifecycle management during dense tensor contractions.
\end{itemize}

A deeper inspection of specific cases, such as the QPE circuit, highlights the impact of our refined criteria to select the operations to store in the \ct. At 30 qubits, the original implementation exhibited a \ct hit ratio of 39\%, whereas the improved framework achieved only 10\%. Despite this lower hit rate, the improved version demonstrates superior temporal performance. This suggests that our selection strategy successfully avoids storing low-impact operations, ensuring that the \ct primarily caches \acp{dd} transformations that provide the highest computational yield per lookup.


Finally, looking at the node counts of each circuit, in most cases, such as AE with 25 qubits, the original version reports $10,940,133$ nodes, while our version reports $23,763,183$. However, despite showing a higher node count, the memory usage of the improved version is noticeably lower, as it occupies only 14 Gb compared to the original’s 99 Gb. This apparent discrepancy is explained by the how the original version counts and uses the nodes: it fails to account for nodes removed from the \ut that remained resident in the \ct, creating a dangling node scenario. While this preserved the functionality of the \ct in some cases, it was structurally unsound and prone to memory corruption or inconsistent results during search operations. Our framework maintains strict node lifecycle management, ensuring that the node counts reflect the actual, coherent state of memory, thereby guaranteeing both performance and numerical correctness.

\begin{table}[]
\begin{tabular}{|cr||r|r|r||r|r|r|}
\hline
\multicolumn{2}{|c||}{Circuit} & \multicolumn{3}{c||}{Original} & \multicolumn{3}{c|}{Improved} \\ \hline
name & \multicolumn{1}{c||}{n} & \multicolumn{1}{c|}{Time (s)} & \multicolumn{1}{c|}{Max\_bck} & \multicolumn{1}{c||}{MeM (GiB)} & \multicolumn{1}{c|}{Time (s)} & \multicolumn{1}{c|}{Max\_bck} & \multicolumn{1}{c|}{MeM (GiB)} \\ \hline  \hline

\multirow{5}{*}{QFT$_\dag$} & 15 & 0.07 & 10 & \textbf{2.70} & 0.07 & \textbf{4} & 2.99 \\
  & 30 & \textbf{24.68} & 3 & 20.80 & 27.58 & \textbf{1} & \textbf{13.08} \\
  & 50 & O.M. & O.M. & O.M. & \textbf{25.06} & \textbf{1} & \textbf{13.10} \\
  & 80 & O.M. & O.M. & O.M. & \textbf{25.69} & \textbf{1} & \textbf{13.11} \\
  & 100 & O.M. & O.M. & O.M. & \textbf{25.80} & \textbf{1} & \textbf{13.12} \\ \hline 
  
\multirow{4}{*}{GHZ$_*$} & 20 & 0.00 & 1 & 3.07 & 0.00 & 1 & \textbf{2.82} \\
  & 30 & 0.00 & 1 & 19.09 & 0.00 & 1 & \textbf{10.82} \\
  & 40 & O.M. & O.M. & O.M. & \textbf{0.00} & \textbf{1} & \textbf{10.81} \\
  & 50 & O.M. & O.M. & O.M. & \textbf{0.01} & \textbf{1} & \textbf{10.80} \\
  \hline
  
\multirow{4}{*}{Graph$_*$} & 20 & 0.00 & 2 & 3.05 & 0.00 & 2 & \textbf{2.81} \\
  & 30 & 0.02 & 1 & 19.07 & 0.02 & 1 & \textbf{10.82} \\
  & 40 & O.M. & O.M. & O.M. & \textbf{0.05} & \textbf{2} & \textbf{10.83} \\
  & 50 & O.M. & O.M. & O.M. & \textbf{0.31} & \textbf{2} & \textbf{10.87} \\
  \hline
  
\multirow{4}{*}{QPE} & 15 & 0.01 & \textbf{2} & 6.25 & \textbf{0.00} & 3 & \textbf{3.06} \\
  & 20 & 0.07 & 2 & 6.37 & \textbf{0.06} & 2 & \textbf{3.06} \\
  & 25 & 115.76  & \textbf{10} & 14.39 & \textbf{93.19} & 13 & \textbf{7.86} \\
  & 30 & 3471.07 & \textbf{10} & 115.21  & \textbf{3,258.76}  & 12 & \textbf{111.77}  \\
  \hline
  
\multirow{4}{*}{QWalk} & 10 & \textbf{2.31} & \textbf{3} & \textbf{2.78} & 2.37 & 50 & 2.90 \\
  & 12 & 44.68 & \textbf{3} & 3.37 & \textbf{44.27} & 23 & \textbf{3.16} \\
  & 14 & 762.17  & \textbf{5} & 4.40 & \textbf{700.60}  & 17 & \textbf{4.19} \\
  & 15 & \textbf{3375.13} & \textbf{6} & 5.79 & 3,415.61  & 14 & \textbf{5.58} \\
  \hline
  
\multirow{2}{*}{RA}  & 15 & \textbf{3.02} & \textbf{7} & 3.55 & 3.11 & 18 & \textbf{2.86} \\
  & 20 & \textbf{187.75}  & \textbf{9} & 21.65 & 215.52  & 19 & \textbf{3.52} \\
  \hline
  
\multirow{4}{*}{AE}
  & 10 & 0.01 & \textbf{9} & 3.00 & 0.01 & 48 & \textbf{2.89} \\
  & 15 & \textbf{0.40} & \textbf{4} & 2.76 & 0.79 & 47 & \textbf{2.92} \\
  & 20 & \textbf{19.89} & \textbf{7} & 5.83 & 26.39 & 14 & \textbf{3.27} \\
  & 25 & \textbf{1017.62} & \textbf{7} & 99.52 & 1,333.01  & 11 & \textbf{14.18} \\
  \hline
  
\multirow{3}{*}{QNN} & 15 & \textbf{1.57} & \textbf{7} & 3.37 & 1.88 & 34 & \textbf{3.09} \\
  & 20 & \textbf{63.50} & \textbf{10} & 12.07 & 95.96 & 35 & \textbf{5.11} \\
  & 25 & \textbf{2790.56} & \textbf{12} & 309.13  & 5,217.02  & 40 & \textbf{67.45} \\
  \hline
  
\multirow{4}{*}{RQC} & 12 & \textbf{4.21} & \textbf{6} & 5.38 & 56.57 & 210  & \textbf{2.96} \\
  & 14 & \textbf{20.64} & \textbf{8} & 9.03 & 210.94  & 128  & \textbf{3.11} \\
  & 16 & \textbf{93.60} & \textbf{9} & 25.64 & 380.79  & 32 & \textbf{3.08} \\
  & 18 & \textbf{465.15}  & \textbf{8} & 104.17  & 1,594.40  & 32 & \textbf{3.40} \\
  \hline

\end{tabular}
  \caption{Comparative of the simulation results between the original and the improved version of the FTDD tool using the \seq contraction plan and the \alp index ordering.}
  \label{tab:comp}
\end{table}



While the previous evaluation isolated the structural impact of our memory management enhancements by utilizing a deterministic and simple baseline (\seq), Table~\ref{tab:spd} details the effect of using two different contraction plan heuristics in our improved \ac{ftdd} framework. Namely, the naive \seq heuristic and the state-of-the-art \cotengra optimization framework. This analysis is critical to determine whether advanced path-optimization strategies, which are traditionally tailored for standard tensor networks, translate into performance gains when paired with \ac{dd}-based quantum simulators.

To prevent redundancy and maintain a focused narrative, Table~\ref{tab:spd} does not include all cases shown in Table~\ref{tab:comp}. Instead, we have strategically selected one or two critical qubit regimes ($n$) per circuit type. These instances represent either the baseline cross-over thresholds where optimization overhead becomes non-negligible, or the maximum scalability limits where one of the two strategies triggers an operational constraint. Performance is quantified using the SpeedUp metric, defined as the execution time of the \seq method divided by that of \cotengra; a value greater than 1 indicates that the advanced path optimizer outperforms the sequential baseline.

At a macroscopic level, the results demonstrate that neither contraction strategy uniformly dominates the other; rather, their efficacy is strictly governed by the structural regularity of the target quantum circuit. The benchmarks can be divided into three contrasting regimes:

\begin{itemize}
    \item \textbf{Regime 1: High Regularity and Scalability Limits (\seq Dominance):} For highly structured topologies like QFT and GHZ with many qubits, the simple \seq baseline outperforms the advanced optimizer. In these cases, \cotengra fails to generate a viable contraction plan within the designated constraints, resulting in an Out-of-Memory ($O.M.$) error during its graph-partitioning phase. Conversely, our improved \ac{ftdd} framework paired with \seq smoothly processes up to 100 qubits for QFT in just 25.80 seconds utilizing a mere 13.12~Gb of RAM (as seen in Table~\ref{tab:comp}). This proves that for highly regular, linear structures, the overhead of advanced path-finding is not only unnecessary but can become a liability that degrades physical scalability.
    
    \item \textbf{Regime 2: Low Regularity and Topological Complexity (\cotengra Dominance):} For circuits characterized by high matrix density, pseudo-randomness, or complex entanglement structures—most notably AE ($n=25$) and RQC ($n=18$)—\cotengra demonstrates a massive, transformative advantage over the \seq heuristic, yielding speedups of 15.77$\times$ and 26.71$\times$, respectively. In these complex topologies, a naive \seq contraction order leads to a explosion in the size of intermediate \acp{dd}. \cotengra successfully mitigates this by finding optimized tree decompositions that keep intermediate representations compact, completely offsetting the temporal overhead of its path-searching phase.

    \item \textbf{Regime 3: Intermediate Regularity, Scalability and Topological Complexity:} In the intermediate regimes (such as QWalk, QNN, RA, and QPE at moderate qubit sizes), the performance of both strategies remains relatively competitive, with speedups hovering closer to 1. As the time employed by \cotengra it is not counted in the overall time, it is most likely that the \seq order must be employed, as the time dedicated by \cotengra to compute an optimal path will be grater than the time saved during the actual contraction process. 
    
\end{itemize}

Then, according to the analysis of the different regimes, the results suggest that an adaptive simulation strategy (relying on simple heuristics for regular structures and invoking advanced graph optimizers exclusively for highly entangled topologies) is the preferred approach for \ac{dd}-based quantum simulation.

\begin{table}[]
\centering
\begin{tabular}{|lr||r|}
\hline
Circuit & qubits & SpeedUp \\ \hline \hline
\multirow{2}{*}{QFT} & 30 & 0.86 \\
 & 100 & O.M. \\
\hline
\multirow{2}{*}{GHZ} & 20 & 2.27 \\
 & 50 & O.M. \\
\hline
Graph$_*$ & 50 & 2.79 \\
QPE & 25 & 1.51 \\
QWalk & 10 & 0.72 \\
RA & 20 & 1.35 \\
AE & 25 & 15.77  \\
QNN & 20 & 0.97 \\
RQC & 18 & 26.71 \\ \hline
\end{tabular}
  \caption{Comparative time speedup between the \seq and \cotengra contraction plan using the improved \ac{ftdd} version. SpeedUp over 1 means \cotengra performs better.}
  \label{tab:spd}
\end{table}

\subsection{Studying the Effect of Table Size}
\label{subsec:rtbl}


The size of the \ut and \ct data structures is a critical factor governing the performance profile of the \ac{ftdd} simulator. Selecting an optimal sizing policy introduces a classic architectural trade-off: oversized tables guarantee sufficient capacity to maximize node reuse and minimize hash collisions, but at the expense of a prohibitive memory footprint. Conversely, overly restrictive capacities trigger frequent node reclamation and a degradation of the effectiveness of the hash tables, preventing the framework from leveraging the structural compression inherent in \acp{tdd}. To systematically evaluate these dynamics, we analyze three distinct table sizing policies under controlled growth boundaries:

\begin{enumerate}
  \item \textit{Fully Exponential Growth} ($2^n (30) - 2^n (30)$): Both tables scale exponentially with the number of qubits n, limited by an upper bound threshold of 30 qubits to prevent unconstrained memory expansion driven solely by table allocation.
  \item \textit{Hybrid Exponential-Fixed} ($2^n (30) - 2^{15}$): This is the default option of the original \ac{ftdd} framework. The \ut scales exponentially (up to n=30, representing a 1\% of the total RAM memory available) while the \ct is fixed at a static capacity of $2^{15}$ entries.
  \item \textit{Static Fixed Allocation} ($2^{30} - 2^{15}$): The size of both tables is kept as $2^{30}$ for \ut and $2^{15}$ for \ct.
\end{enumerate}


Table~\ref{tab:tab} presents a comprehensive performance comparison across these allocation regimes using the \seq contraction plan. Empirical observations reveal that the Static Fixed policy generally yields the best execution time for the majority of the evaluated benchmarks. However, notable exceptions exist; for instance, the QNN circuit at 25 qubits exhibits degraded temporal performance under larger table configurations.

This behavior highlights a critical trade-off within our garbage collection mechanism. In this specific case, expanding the table capacity only slightly reduces the average bucket density (e.g., from 5 to 3 nodes per bucket). While this marginal reduction accelerates individual lookup times, it is vastly outweighed by the unproductive table traversal overhead imposed on the garbage collector. Because the total number of buckets scales up, each invocation of the garbage collection routine must scan a substantially larger memory space to identify inactive nodes. Consequently, larger table sizes only become computationally advantageous when the capacity expansion induces a major reduction in collision density (e.g., dropping from 20 nodes per bucket down to 5), thereby offsetting the sequential traversal overhead.

Surprisingly, the Fully Exponential policy proves to be worse in both spatial and temporal metrics, whereas the Hybrid configuration emerges as the most efficient approach regarding memory footprint optimization. This phenomenon stems from a dual-layer bottleneck. When the \ct is too small, the simulator fails to cache and retrieve intermediate operations, leading to redundant computations and severe time degradation. Conversely, an oversized \ct forces the system to retain a massive volume of structural nodes to support those cached operations. If these operations are not frequently reused, the memory overhead escalates rapidly, which in turn degrades garbage collection efficiency and impairs both execution time and memory limits.


Finally, it is important to underscore that for circuits scaling beyond 30 qubits, the Hybrid and Static Fixed policies converge to identical structural dimensions. Consequently, both paradigms represent optimal execution strategies for large-scale, high-performance regimes where \ac{ftdd} demonstrates its greatest scalability, such as in QFT, GHZ, and Graph benchmarks.

\begin{table}[]
\begin{tabular}{|cr||rr||rr||rr|}
 \hline
 \multicolumn{2}{|c||}{} & \multicolumn{2}{c||}{Hybrid Exp-Fix} & \multicolumn{2}{c||}{Fully Exp. Growth} & \multicolumn{2}{c|}{Static Fix. Alloc.} \\
\multicolumn{2}{|c||}{Circuit} & \multicolumn{2}{c||}{$2^n (30) - 2^{15}$} & \multicolumn{2}{c||}{$2^n (30) - 2^n (30)$} & \multicolumn{2}{c|}{$2^{30} - 2^{15}$} \\
name & \multicolumn{1}{c||}{n} & \multicolumn{1}{c}{Time (s)} & \multicolumn{1}{c||}{M (GiB)} & \multicolumn{1}{c}{Time (s)} & \multicolumn{1}{c||}{M (GiB)} & \multicolumn{1}{c}{Time (s)} & \multicolumn{1}{c|}{M (GiB)} \\ \hline \hline

\multirow{4}{*}{AE}
  & 10 & 0.01 & \textbf{2.89} & 0.01 & 3.12 & 0.01 & 10.95  \\
  & 15 & 0.79 & \textbf{2.92} & 0.58 & 3.15 & \textbf{0.42} & 11.05  \\
  & 20 & 26.39  & \textbf{3.27} & 69.77  & 4.60 & \textbf{17.93}  & 14.07  \\
  & 25 & 1,333.01 & \textbf{14.18}  & 3,476.56 & 55.27  & \textbf{1,044.67} & 106.00 \\ \hline
  
\multirow{4}{*}{GHZ$_*$} & 20 & 0.00 & \textbf{2.82} & 0.00 & 3.13 & 0.00 & 11.00  \\
  & 30 & 0.00 & \textbf{10.82}  & 0.00 & 19.14  & 0.00 & 10.99  \\
  & 40 & 0.00 & \textbf{10.81}  & 0.00 & 19.15  & 0.00 & 10.99  \\
  & 50 & 0.01 & \textbf{10.80}  & 0.01 & 19.17  & 0.01 & 10.98  \\ \hline
  
\multirow{4}{*}{Graph$_*$} & 20 & 0.00 & \textbf{2.81} & 0.00 & 3.18 & 0.00 & 10.99  \\
  & 30 & \textbf{0.02} & \textbf{10.82}  & 0.03 & 19.18  & \textbf{0.02} & 11.00  \\
  & 40 & \textbf{0.05} & \textbf{10.83}  & 0.08 & 19.17  & \textbf{0.05} & 11.01  \\
  & 50 & 0.31 & \textbf{10.87}  & 0.41 & 19.31  & \textbf{0.30} & 11.04  \\ \hline
  
\multirow{3}{*}{QNN} & 15 & 1.88 & \textbf{3.09} & 1.89 & 3.17 & \textbf{1.67} & 11.25  \\
  & 20 & 95.96  & \textbf{5.11} & 158.95 & 6.38 & \textbf{60.90}  & 19.94  \\
  & 25 & \textbf{5,217.02} & \textbf{67.45}  & T.O. & T.O. & 5,789.79 & 157.04 \\ \hline
  
\multirow{4}{*}{QPE} & 15 & 0.00 & \textbf{3.06} & 0.00 & 3.15 & 0.00 & 11.02  \\
  & 20 & \textbf{0.06} & \textbf{3.06} & 0.07 & 3.14 & \textbf{0.06} & 11.01  \\
  & 25 & 93.19  & \textbf{7.86} & 107.50 & 13.97  & \textbf{79.75}  & 18.76  \\
  & 30 & 3,258.76 & 111.77 & T.O. & T.O. & \textbf{3,258.43} & \textbf{111.67} \\ \hline
  
\multirow{4}{*}{QWalk} & 10 & 2.37 & \textbf{2.90} & \textbf{2.26} & 3.18 & \textbf{2.26} & 11.06  \\
  & 12 & 44.27  & \textbf{3.16} & 46.08  & 3.44 & \textbf{42.97}  & 11.32  \\
  & 14 & \textbf{700.60} & \textbf{4.19} & 703.39 & 4.47 & 741.54 & 12.37  \\
  & 15 & 3,415.61 & \textbf{5.58} & 3,482.46 & 5.86 & \textbf{3,344.70} & 13.77  \\ \hline
  
\multirow{2}{*}{RA}  & 15 & 3.11 & \textbf{2.86} & \textbf{2.99} & 3.17 & 3.11 & 11.44  \\
  & 20 & 215.52 & \textbf{3.52} & 311.30 & 4.30 & \textbf{175.44} & 25.40  \\ \hline
  
\multirow{4}{*}{RQC} & 12 & 56.57  & \textbf{2.96} & 15.15  & 3.16 & \textbf{5.43} & 12.08  \\
  & 14 & 210.94 & \textbf{3.11} & 89.20  & 3.26 & \textbf{24.60}  & 15.76  \\
  & 16 & 380.79 & \textbf{3.08} & 642.84 & 3.50 & \textbf{118.05} & 32.39  \\
  & 18 & 1,594.40 & \textbf{3.40} & 5,126.57 & 5.44 & \textbf{831.56} & 110.94 \\ \hline
  
\multirow{5}{*}{QFT$_\dag$} & 15 & \textbf{0.07} & \textbf{2.99} & \textbf{0.07} & 3.07 & 0.09 & 10.98  \\
  & 30 & 27.58  & \textbf{13.08}  & \textbf{23.96}  & 23.93  & 27.56  & \textbf{13.08}  \\
  & 50 & 25.06  & \textbf{13.10}  & \textbf{21.20}  & 23.94  & 25.06  & \textbf{13.10}  \\
  & 80 & 25.69  & \textbf{13.11}  & \textbf{21.53}  & 23.97  & 25.40  & \textbf{13.11}  \\
  & 100 & 25.80  & \textbf{13.12}  & \textbf{21.70}  & 24.04  & 25.47  & \textbf{13.12} \\ \hline  
\end{tabular}
  \caption{Comparative of the simulation results between different table size for the \ut (left) and \ct (right) with the FTDD tool using the \seq contraction ordering and the \alp index ordering.}
  \label{tab:tab}
\end{table}

\subsection{Evaluation of Index Ordering Strategies}
\label{subsec:order}

The \alp ordering of indices within \ac{dd}-based \ac{tn} simulation represents a critical factor in the overall performance of the tool, yet its implications are often overlooked. In quantum circuit simulation, the choice of index ordering dictates the internal variable ordering of the underlying \acp{dd}, directly governing the structural compression of the state vector and the efficiency of operations during contraction operations. A sub-optimal ordering can lead to exponential node duplication and severe cache thrashing. Consequently, systematically evaluating and understanding the impact of different ordering strategies is essential to improve the performance of the simulator.

\begin{figure}[ht]
  \centering
  \includegraphics[width=\linewidth]{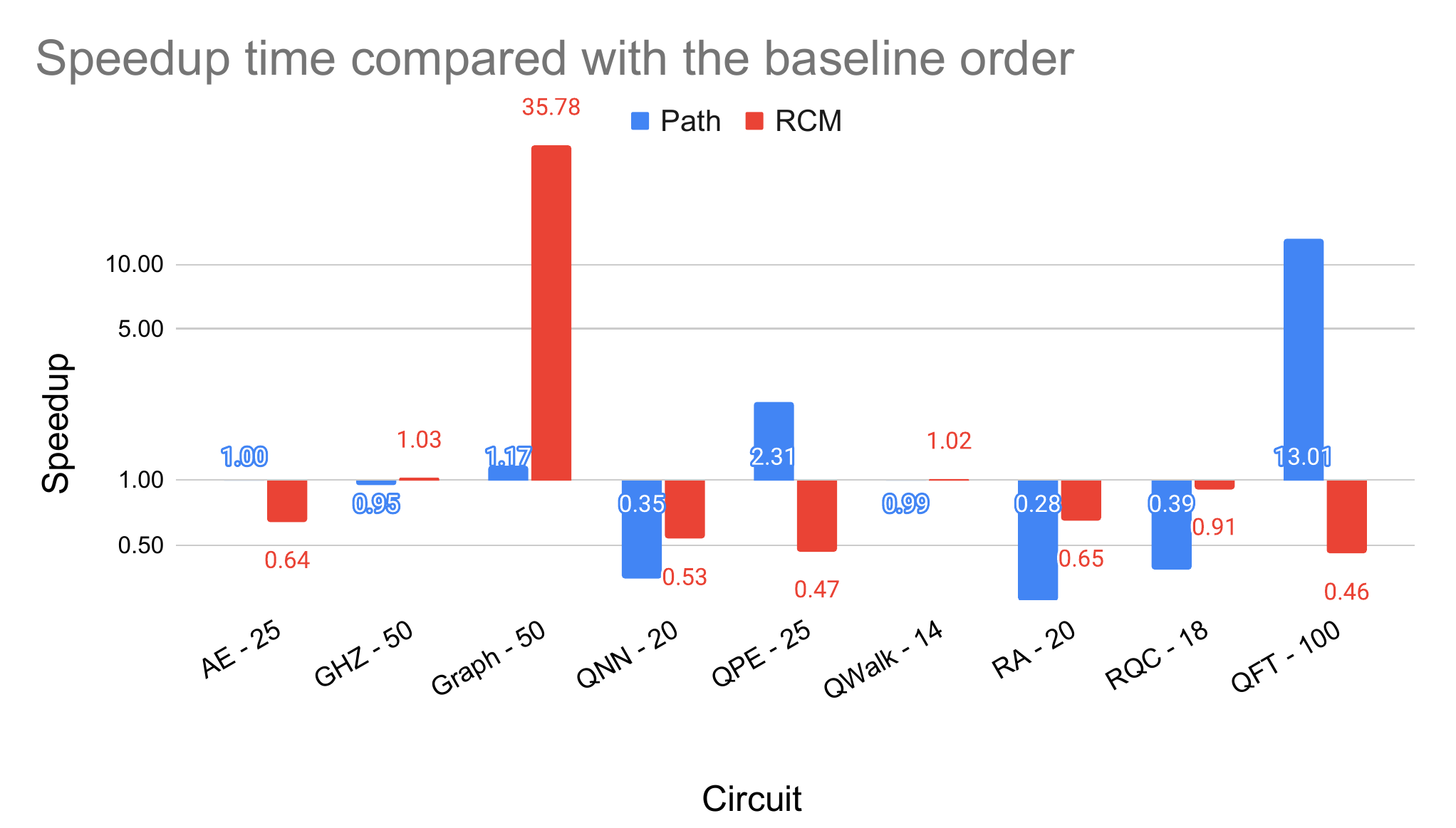}
  \caption{Performance speed-up comparison of the \mpath and \rcm index ordering methods relative to the \alp baseline, evaluated using the \ac{ftdd} tool under the \seq contraction strategy.}
  \label{fig:spd}
\end{figure}

Figure~\ref{fig:spd} illustrates the comparative performance of the baseline \alp index ordering against our proposed \rcm and \mpath strategies and using the \seq contraction plan. The experimental results reveal that the effectiveness of each ordering method is highly sensitive to the underlying structure of the quantum circuit. For benchmark circuits such as AE, GHZ, and QWalk, performance differentials across the three methods remain marginal (below $2\times$). Conversely, specific circuit topologies show significant sensitivity: for the Graph circuit, the \rcm strategy achieves a speed-up exceeding $35\times$, while for QFT, the \mpath ordering yields a speed-up surpassing $13\times$. These findings underscore the fact that no single index ordering provides a universally optimal performance gain; rather, the selection of an appropriate strategy is a critical factor in maximizing the tool's overall computational throughput.

\section{Conclusions}
\label{sec:con}

This work presents a series of systematic architectural and memory-management enhancements designed to improve the memory scalability of the \ac{ftdd} quantum circuit simulator. By transitioning from an unbounded dynamic allocation model to a hardware-aware, fixed-footprint framework utilizing strict node lifecycle tracking and refined cache filtration, we address the chronic memory vulnerability that typically compromises \ac{dd}-based \ac{tn} contractions. 

The experimental evaluation provides a realistic view of the inherent trade-offs involved in these optimizations. On one hand, the proposed framework successfully controls memory inflation, enabling the simulation of highly regular topologies (such as QFT) up to 100 qubits within a stable, bounded memory and time consumption. On the other hand, we acknowledge that our strict node tracking introduces a non-negligible computational and administrative overhead. In high-depth, highly entangled regimes (such as RQC and QNN), this overhead manifests as a significant temporal penalty, where the execution time cost of maintaining structural correctness and strict node coherence outweighs the immediate gains in memory stability. But, if we do not restrict the execution time, the improved version could simulate even this kind of circuits with a larger number of qubits.

This delicate balance is further highlighted by our investigation into table sizing policies, which revealed a critical and counter-intuitive interplay between memory capacity and our garbage collection mechanism. We observe that expanding the capacities of the \ut and \ct does not act as a universal remedy. While a Static Fixed allocation generally minimizes individual lookup times by reducing bucket density, it can inadvertently bottleneck performance in specific variational circuits like QNN. In these cases, minor reductions in hash collisions are entirely eclipsed by the severe sequential traversal overhead imposed on the garbage collector  (which must scan an expanded memory space to identify and reclaim inactive nodes) and the \ut (which must scan through an expanded linked list to lookup for the nodes). Furthermore, our analysis indicates that an oversized \ct forces the system to retain an excessive volume of structural nodes to support cached operations, rapidly degrading efficiency unless the operations are frequently reused. Consequently, the Hybrid Exponential-Fixed configuration emerges as the most efficient approach for memory footprint optimization, while both the Hybrid and Static Fixed paradigms successfully converge to provide optimal execution strategies for large-scale, high-performance regimes exceeding 30 qubits.

Furthermore, our exploration into contraction paradigms and internal variable layouts demonstrates that the optimal configuration of a \ac{dd}-based simulator is intimately tied to the topological features of the target quantum algorithm. Advanced path optimizers like \cotengra offer transformative temporal advantages for complex, low-regularity networks, yet they can become resource liabilities in highly structured linear circuits where simpler sequential heuristics suffice. Similarly, index ordering strategies like \rcm and our new \mpath heuristic are not silver bullets; their capacity to accelerate lookups and minimize node duplication is highly localized to specific circuit families.

As part of our future research, our aim is to extend the evaluation of the proposed tool in two main directions. First, we intend to modify the internal structures to bypass their 100 qubit limit. The objective of this evaluation is to identify the system's operational ceiling without the hard limit of the internal structures and establish precise performance bounds, thereby clarifying the scalability limits under extreme load conditions.

Second, we will address the classification of the three index orders presented in this study. Our current findings indicate that distinguishing between these orders is a non-trivial task, as we have been unable to establish a direct correlation between circuit topology and performance metrics through traditional analysis. Consequently, we aim to develop an AI-based model designed to automatically select the best index ordering. Using machine learning, we anticipate uncovering the underlying patterns and latent relationships between circuit structure and performance that remain elusive to conventional methods.

\section*{Declarations}
\begin{itemize}
\item \textbf{Funding} This work has been funded by project UJI-B2021-26 of the Universitat Jaume I and by grant PID2020-113656RB-C21 funded by MCIN/AEI/10.13039/50110011033 and MICINN.
\item \textbf{Conflict of interest} The authors have no conflicts of interest to declare that are relevant to the content of this article.
\item \textbf{Ethics approval} Not applicable.
\item \textbf{Consent to participate} Not applicable.
\item \textbf{Consent for publication} All authors read and approved the final manuscript.
\item \textbf{Availability of data and materials} The circuits used and the executable scripts are available along with the code.
\item \textbf{Code availability} https://github.com/voliva-esp/FETDD
\item \textbf{Authors’ contributions} V.L.O. implemented the codes and executed all the experiments on the computing platforms. J.M.B. and M.C. supervised the work and helped to conceive the codes and experiments. All authors discussed the results and contributed to the final version of the manuscript.
\end{itemize}

\bibliography{sn-bibliography}

\end{document}